%% file: paper.tex
\documentclass[oribibl]{llncs}
\usepackage{llncsdoc}

\usepackage{graphicx}
\usepackage{amssymb}
\usepackage{amsmath}

\newcommand{\ket}[1]{|#1\rangle}
\newcommand{\bra}[1]{\langle#1|}
\newcommand{\braket}[2]{\langle#1|#2\rangle}
\newcommand{\calH}{{\cal H}}

\newcommand{\Tr}{\mathrm{Tr}}

\author{William Blacoe}
\institute{School of Informatics, University of Edinburgh,  Scotland\\\email{w.b.blacoe@sms.ed.ac.uk}}
\title{On Quantum Generalizations of\\Information-Theoretic Measures and\\their Contribution to Distributional Semantics}

\begin{document}

\maketitle

\begin{abstract}

Information-theoretic measures such as relative entropy and correlation are extremely useful when modeling or analyzing the interaction of probabilistic systems. We survey the quantum generalization of 5 such measures and point out some of their commonalities and interpretations. In particular we find the application of information theory to distributional semantics useful. By modeling the distributional meaning of words as density operators rather than vectors, more of their semantic structure may be exploited. Furthermore, properties of and interactions between words such as ambiguity, similarity and entailment can be simulated more richly and intuitively when using methods from quantum information theory.

\end{abstract}

\input{intro}
\input{general}
\input{linguistics}
\input{discussion}

\bibstyle{splncs}
\bibliographystyle{splncs}
\bibliography{references}

\end{document}

%% file: intro.tex
\section{Introduction}\label{sec:intro}

The notion of representing quantum states as density operators, rather than as kets, was introduced by von Neumann \cite{Neumann1927}. This transition brought with it an increased expressivity and complexity. While a ket can be understood to be a (joint) probability distribution, a density operator is a probability distribution over eigen kets.
This additional dimension of structure ought to be exploited by various fields which involve uncertainty and interaction between multiple systems. Many of the measures developed in probability and information are applicable to density operators. Since a density operator is a distribution over kets rather than atomic symbols, these measures have been extended to take advantage of the added structure contained in the kets.

One field in particular that promises to profit from quantum information theory is distributional semantics. This approach to meaning in human language makes use of Firth's hypothesis that ``a word is characterized by the company it keeps'' \cite{firth1957}. Much work in computational linguistics has evolved from this hypothesis, in which the meanings of words are typically encoded in vectors \cite{Turney:Pantel:2010,Erk:Survey:2012}. However,
density operators have been used to model linguistic meaning in several existing works. Sordoni and Nie \cite{Sordoni2014} perform information retrieval by encoding two types of a word's context in a density operator. They leverage the strengths of language models and vector space models in a combined data structure. Balkir \cite{Esma2014} generalizes the use of word vectors to density operators to take advantage of its probabilistic interpretation. Piedeleu et al. \cite{Robin2015} extend the tensor framework of \cite{Coecke2010} to apply to density operators. The result is a powerful theoretical model that compositionally derives the distributional meaning of a sentence, given its syntax tree and the lexical density operators.
Blacoe \cite{blacoe-kashefi-lapata:2013:NAACL-HLT} have another approach to representing the meaning of words and sentences. In their framework the density matrices for all words, phrases and sentences live in the same semantic space. This makes the derivation of rich sentence representations and their information-theoretical analysis tractable.

In Section 2 we survey several information-theoretic measures and mathematically examine the differences between them and their quantum analogues. In Section 3 we then observe how density matrices used in the field of distributional semantics have made use of quantum information theory. We also provide intuitions about how the quantum generalizations of information-theoretic measures is advantageous to modeling human language processes. Section 4 concludes the paper.

%% file: general.tex
\section{Quantum Generalizations}\label{sec:general}

In this section we compare aspects of classical probability theory with their quantum analogue. Generalising probability distributions to density operators has many advantages. While probability distributions are generally over atomic symbols, a density operator can be considered a probability distribution over eigen kets. Depending on the kets' internal structure, they can encode a lot of information about some system. If the system is a complex of more than one subsystem, the kets have order $>1$ and may encode correlation among the subsystems. The dimensions of the Hilbert spaces for each subsystem also allow for holding much information. Each eigen ket can therefore represent a joint probability distribution. If its values are complex values, of course we have an additional degree of freedom to represent information.

Given the complexity of density operators, it is difficult to grasp the interactions among all of the information they contain. This section will illuminate aspects of probability distributions, classical and quantum, by mathematically analyzing measures from probability theory and information theory. The quantum generalization of these measures has had much impact in the field of quantum information theory \cite{Nielsen:Chuang:10}. We compare several such measures and find a pattern among their quantum generalizations. This will help us build intuitions about the gain that can come from rising to the quantum level.

Throughout this section we will be using random variables $X$ and $Y$. More formally, let $X$ be a random variable with possible outcomes $x_1, ..., x_n$ and corresponding probabilities $p(x_1), ..., p(x_n)$. $Y$ has possible outcomes $y_1, ..., y_m$ with corresponding probabilities $p(y_1), ..., p(y_m)$.
Analogously to them we will use density matrices $A$ and $B$. Let their respective spectral decomposition be $A = \sum_{i=1}^n \alpha_i \ket{a_i}\bra{a_i}$ and $B = \sum_{j=1}^m \beta_j \ket{b_j}\bra{b_j}$. This means that the $\alpha_i$ are real, positive values that sum to 1, and the same goes for the $\beta_j$. Hence, they are the probabilities over their corresponding kets of unit norm $\ket{a_i}$ and $\ket{b_j}$.

Two functions of relevance here are $\sqrt{\cdot}$ and $\log(\cdot)$. When applied to a density operator $A$ they ``transfer'' to the operator's eigen values. The unique positive square root of $A$ is:
\begin{gather}
\sqrt{A} = \sum_i \sqrt{\alpha_i} \ket{a_i}\bra{a_i}
\end{gather}
and the logarithm of $A$ is:
\begin{gather}
\log A = \sum_i \log(\alpha_i) \ket{a_i}\bra{a_i}
\end{gather}

\subsection{Von Neumann Entropy}

One fundamental quantum generalization is that of entropy. While the Shannon entropy for a random variable $X$ is $H(X) = \sum_i p(x_i) \log_2 p(x_i)$, its quantum analogue is von Neumann entropy \cite{Nielsen:Chuang:10}:
\begin{gather}
S(A) = \Tr(A\,\log_2 A) = \sum_i \alpha_i \log_2 \alpha_i
\end{gather}
Quantum entropy underlies several of the information-theoretic measures that this section addresses. It is a measure of the information content of a quantum state. Whereas the entropy function is defined for only one probability distribution, the following measures will be over pairs of probability distributions.

\subsection{Incompatibility}

At the heart of the difference between classical and quantum probability lies the matter of incompatibility. Two systems are incompatible whenever they are correlated and can therefore not be treated independently. Mathematically, this is the case when the systems' operators $A, B$ do not commute, that is $AB\neq BA$. Another way of making this obvious is to compare $A$'s and $B$'s eigen bases.
If there is some one-to-one alignment between $A$'s and $B$'s eigen kets such that $\ket{a_k} = \ket{b_k}$ for $k = 1, ..., \mathrm{min}(n, m)$ then we do not have a quantum effect. In this case the interaction of $A$ and $B$ reduces to the classical case where $\{\alpha_i\}_i$ and $\{\beta_j\}_j$ are regular probability distributions.
However, if $A$'s and $B$'s eigen bases do not match up so neatly, then non-classical effects take place. In the following we will analyze how well-known measures for probability distributions are effected by the presence of this incompatibility.

%
%
%

\subsection{Measurement}\label{sec:measurement}



In the classical paradigm, a system's properties are independent of the way in which they are observed. So if there is uncertainty about the state of a system, that uncertainty is not due to any aspect of the system. It is merely epistemic, that is, in the observer's mind. We represent this uncertainty with the random variable $X$. Let its possible states $x_1, ..., x_n$ be real values. Then the overall value of $X$, given the observer's uncertainty, is its expected value:
\begin{gather}
\mathbb{E}(X) = \sum_i x_i \times p(x_i)
\end{gather}

Quantum mechanics questions the realist perspective mentioned above and posits that uncertainty can also be ontological. That is to say,
there are propositions which have an uncertain value, independent of our (lack of) knowledge thereof. Let us represent our knowledge of the observed system by $A$ and the mixed state of the system as $B$. We cast the situation of observing the system, given our limited knowledge, as a measurement scenario. Thus, $A$ is regarded as the observable for measuring $B$, and the projective measures $A_i$ come from $A$'s eigen kets: $A_i = \ket{a_i}\bra{a_i}$. The measurement's outcome value will be one of $A$'s eigen values.

However, we will not allow the measurement process to collapse. Von Neumann \cite{vonNeumann:1932} splits the process into two steps: (1) measurement, which assigns each possible outcome state $A_i$ a probability $p_i$, and (2) observation, which collapses the system's outcome state to $A_i$ with probability $p_i$. If the second step never occurs, then the system is furthermore in a mixed state, also called statistical state, which is the sum of all $A_i$ weighted by the corresponding probabilities $p_i$. In a projective measurement of system $B$ the probability $p_i$ for measuring $A$'s eigen value $\alpha_i$ is:
\begin{gather}
p_i = \bra{a_i} B \ket{a_i} = \bra{a_i}\left(\sum_j \beta_j \ket{b_j}\bra{b_j}\right)\ket{a_i} = \sum_j \beta_j \braket{a_i}{b_j}\braket{b_j}{a_i}
\end{gather}
Therefore, if we forego the collapse, the statistical outcome value is
\begin{gather}
\sum_i \alpha_i \times p_i = \sum_i \alpha_i \left(\sum_j\beta_j \braket{a_i}{b_j}\braket{b_j}{a_i}\right) = \sum_{i,j} \alpha_i\beta_j \braket{a_i}{b_j}\braket{b_j}{a_i}
\end{gather}

\subsection{Fidelity}\label{sec:fidelity}
The Bhattacharyya coefficient is a useful means to compute the similarity of two probability distributions. It is defined as:
\begin{equation}
BC(X, Y) = \sum_{i} \sqrt{p(x_i) p(y_i)}
\end{equation}
Its minimum value is 0, and its maximum value 1 is reached if and only if $X=Y$. For two probability distributions to be equal they must have the same amount of probabilities and the same values in the same position (marked by the index $i$). Any permutation of values in either distribution will lead to the Bhattacharyya coefficient to be less than 1.

The quantum analogue is the fidelity of two quantum states. This is a common measure for measuring the similarity between two mixed states. It is defined as:
\begin{gather}
F(A, B) = \Tr(\sqrt{A} \sqrt{B}) = \sum_{i,j} \sqrt{\alpha_i \beta_j}\ \braket{a_i}{b_j}\braket{b_j}{a_i}\\
= \sum_i \sqrt{\alpha_i} \left(\sum_j \sqrt{\beta_j}\,\braket{a_i}{b_j}\braket{b_j}{a_i}\right) \nonumber
\end{gather}
Again its value ranges from 0 to 1, with $F(A, B) = 1$ if and only if $A=B$. So, for this to happen, not only the eigen values need to align, but also $A$'s and $B$'s eigen kets. This is obviously a classical case, as $A$ and $B$ must be diagonal in the same base in order to be equal.

An important difference between $BC$ and $F$ is how the probabilities align. In the classical version, the closer the probabilities with the same index are to each other, the higher the Bhattacharyya coefficient. Fidelity makes no requirements about indices being equal. As long as there is some one-to-one connection between $A$'s and $B$'s eigen kets via their inner product, the corresponding eigen values' square roots will be multiplied and then summed over.


\subsection{Relative Entropy}\label{sec:relativeEntropy}
Another function over probability distributions is the Kullback-Leibler (KL) divergence. It measures how far one probability distribution diverges from another. Let a given system's true probabilistic state be $X$, and let it be approximated by a distribution $Y$. The divergence $Y$ from $X$ can be thought of as the amount of additional information required to correct the errors made by this approximation. The KL divergence of $Y$ from $X$ is defined as:
\begin{gather}
D_{KL}(X||Y) = \sum_i p(x_i) \log_2 p(x_i) - \sum_i p(x_i) \log_2 p(y_i) = \sum_i p(x_i) \log_2 \frac{p(x_i)}{p(y_i)}
\end{gather}
It is clear that this measure is not symmetric. Furthermore, it is closely tied to the entropy measure.
As is the case with entropy, KL divergence lends itself well to quantum generalization \cite{Vedral:RelativeEntropy}. How much information is lost when describing a quantum state $A$ using an approximated state representation $B$? Again, this measure is related to entropy, and uses the same letter $S$:
\begin{gather}
S(A||B) = \Tr(A\ \log_2 A) - \Tr(A\ \log_2 B)\\
= \sum_i \alpha_i \log_2 \alpha_i - \sum_{i,j} \alpha_i \log_2(\beta_j)\,\braket{a_i}{b_j}\braket{b_j}{a_i} \nonumber\\
= \sum_i \alpha_i \log_2 \alpha_i - \sum_i \alpha_i \left(\sum_j \log_2(\beta_j)\,\braket{a_i}{b_j}\braket{b_j}{a_i}\right) \nonumber 
\end{gather}
Relative entropy is often used intuitively as a measure of distance between the states $A$ and $B$. We are guaranteed by Jensen's inequality \cite{Jensen:Inequality} that it is always non-negative. It is, however, not a metric since it is not even symmetric in its arguments. We see that it extends the KL divergence by bringing together all pairs of $A$'s and $B$'s eigen kets in an inner product.

\subsection{Common Theme}
Measurement, fidelity and relative entropy are information-theoretic measures involving (at least) two mixed operators. We observe that their extension of their classical analogue for probability distributions follows a pattern. It involves the term $\braket{a_i}{b_j}\braket{b_j}{a_i}$, a contribution made possible by using density operators as representations. But what does this ``quantum term'' contribute? What use is this extension to us? We will identify theoretical advantages here, and in the following section apply them to the field of distributional semantics.

As mentioned earlier, the quantum term has something to do with the alignment and interaction of probabilities. In the classical paradigm, the alignment is straightforward. That is, the indices $i$ and $j$ collapse to one index, and the probabilities' alignment is explicitly defined. We also mentioned that this one-to-one alignment is simulated by density operators whenever they commute, because then for each pair of eigen kets $\ket{a_i}, \ket{b_j}$ the inner product is either 0 or 1. Whenever $\braket{a_i}{b_j} = 1$ the corresponding probabilities $\alpha_i$ and $\beta_j$ interact.

The interaction of the density operators' eigen values uses varying functions for the different measures: In Section \ref{sec:measurement} they get multiplied, in Section \ref{sec:fidelity} their square roots get multiplied, and in Section \ref{sec:relativeEntropy} $\alpha_i$ gets multiplied with $\log_2 \beta_i$. This is true for the classical case as well as for the quantum case. However, when $A$ and $B$ do not commute, the value $\beta_i$ is amended by the quantum term in a systematic way:

In Section \ref{sec:measurement} if $A$ and $B$ are incompatible the product $\alpha_i \times \beta_i$ generalizes as follows:
\begin{gather}
\beta_i \mapsto \sum_j \beta_j \braket{a_i}{b_j}\braket{b_j}{a_i}
\end{gather}
In Section \ref{sec:fidelity} the product of square roots $\sqrt{\alpha_i} \times \sqrt{\beta_i}$ is affected thus:
\begin{gather}
\sqrt{\beta_i} \mapsto \sum_j \sqrt{\beta_j} \braket{a_i}{b_j}\braket{b_j}{a_i}
\end{gather}
And in Section \ref{sec:relativeEntropy} the term $\alpha_i \times \log_2(\beta_i)$ undergoes this change:
\begin{gather}
\log_2(\beta_i) \mapsto \sum_j \log_2(\beta_j) \braket{a_i}{b_j}\braket{b_j}{a_i}
\end{gather}

Each generalization has in common that the function applied to $\beta_i$ is now applied to all $\beta_j$ which is then weighted by the quantum term. What is gained by this added term? How does this change the interaction between $A$'s and $B$'s eigen values? We might say that the alignment of $\{\alpha_i\}_i$ and $\{\beta_j\}_j$ has been very much relaxed. Due to the quantum term, all eigen kets interact, making it possible for all eigen values to interact. There is no longer a strict one-to-one correspondence of probabilities. One great advantage is that when designing density operators for states one does not need to manually pair up the states' possible outcomes, or whatever it is the eigen kets encode. Cross-overs or ``soft alignments'' are now possible, making the interaction of the two distributions richer and more versatile.

\subsection{Entanglement}
We mention one further quantum generalization here: The classical correlation among (random) variables has an analogue called quantum correlation or entanglement.

Let $p(x_i, y_j)$ be the joint distribution that jointly represents the distributions from variables $X$ and $Y$. If the two variables are correlated then this is measurable by their mutual information, which is defined as:
\begin{equation}
I(X; Y) = H(X) + H(Y) - H(X, Y) = \sum_{i,j} p(x_i, y_j) \log_2 \frac{p(x_i, y_j)}{p(x_i) p(y_j)}
\label{eq:mutualInformation}
\end{equation}
If $p(x_i, y_j) = p(x_i) \times p(y_j)$ for all $i,j$ then $p(x_i, y_j)$ is merely a ``product distribution''. In this case there is no correlation among $X$ and $Y$ and the mutual information is 0 (the logarithm's value in Equation \ref{eq:mutualInformation} vanishes). However, if $X$ and $Y$ are in the least correlated, then $I(X;Y)$ is positive.

The matter of correlation in quantum information theory has been the topic of much research \cite{Vedral:discord}. It turns out that quantum correlation does not subsume classical correlation. Rather in different quantum states they may or may not coexist. In what follows we follow Werner's \cite{Werner1989} differentiation of the two different species of correlation.

As with entropy, formulating a quantum analogue of mutual information is straightforward. If $C$ is the density operator that describes a bipartite system in $\calH_1\otimes\calH_2$, then the correlation among its two subsystems can be measured by $C$'s quantum mutual information. Let $A$ and $B$ be reductions of $C$ via the partial trace: $A = Tr_{\calH_2}(C)$ and $B = Tr_{\calH_1}(C)$. If $C = A \otimes B$ is the product state of $A$ and $B$ then the subsystems are in no way correlated, that is they are independent.

If $C$ is not a product state, it may nevertheless be seperable. If this is the case then there are probabilities $\{\gamma_i\}_i$ and matrices $\{A_i\}_i$ and $\{B_i\}_i$ such that
\begin{gather}
C = \sum_i \gamma_i A_i \otimes B_i
\end{gather}
In such a situation it is possible for $C$ to contain classical correlation, but no quantum correlation.

Beyond independence and separability, another possibility is entanglement, which $C$ contains if it is outside of the set of all separable sets. Entangled states may still contain classical correlation. \cite{Vedral:Correlation} use relative entropy to make clear the difference between different kinds of (possibly simultaneous) correlation: They call \textit{total correlation} the sum of classical correlation and quantum correlation. This is defined analogously to classical mutual information using the function for von Neumann entropy:
\begin{gather}
Corr_{total} = S(A) + S(B) - S(C) = S(C || A\otimes B)
\end{gather}
This indicates that the total correlation between subsystems $A$ and $B$ can be expressed in terms of relative entropy. The information that the two systems share in their combined state $C$ is the amount of information lost when approximating them as the product of their reduced states. In other words, their correlation is the ``distance'' between their current state and their totally uncorrelated state.

\cite{Vedral:Correlation} divide the total correlation into two parts using the set $D = \{D_m\}_m$ of all separable states. Entanglement is the additional amount of correlation found in states for being outside of $D$. Therefore $C$'s entanglement can be quantified as its distance to the set $D$, which is the distance of $C$ to the ``nearest'' separable state using relative entropy as distance measure:
\begin{gather}
Corr_{quantum} = \min_{D_m\in D} S(C || D_m)
\end{gather}




The classical correlation in $C$ is whatever is left over from the total correlation:
\begin{gather}
Corr_{classical} = Corr_{total} - Corr_{quantum}
\end{gather}

As before, the additional correlation among quantum subsystems is made possible through representing them as mixed states. Due to the interaction among eigen kets, the density operators' eigen values can interact in a way that goes beyond the correlation among two regular probability distributions.

%% file: linguistics.tex
\section{Applicability in Linguistics}\label{sec:linguistics}



This section goes through the functions mentioned in Section \ref{sec:general} and applies them to distributional semantics. The works mentioned in Section \ref{sec:intro} which use density operators for linguistic tasks use these functions to compute values relevant to natural language technology, such as the degree of similarity and entailment. First we review an example of how a density operator representing a word's meaning is derived from a corpus of text. Then we use the ambiguous example words \textit{book} and \textit{schedule} to illustrate their interpretation and amenability to quantum information theory.

\subsection{An Example}\label{sec:example}
The following is an example of generating lexical density matrices in the style of \cite{blacoe-kashefi-lapata:2013:NAACL-HLT}. They scower through a large corpus of lemmatized and dependency-parsed text documents, counting for each target word how often it occurs with certain contexts. In distributional semantics it is common to encode the (relative) frequency of context words near the target word in a vector. However, \cite{blacoe-kashefi-lapata:2013:NAACL-HLT} define contexts to be syntactical neighborhoods, that is the set of all words in the sentence connected directly to the target word via some dependency relation. For each type of relation $R_i$ there is a vector space $V_{R_i} = \mathbb{C}^d$ whose dimensions encode the $d$ most common words under those relations. The overall Hilbert space is then the tensor product $\calH = V_{R_1} \otimes ... \otimes V_{R_n}$. A ket $\ket{w_i, doc_j}\in\calH$ represents the (relative) frequencies of the word $w_i$ occurring with common dependency neighborhoods in document $doc_j$ of the corpus.

Density matrices $\widehat{w_i}$ are created from these ``document kets'' as follows:
\begin{gather}
\widehat{w_i} = \frac{\sum_j \ket{w_i, doc_j}\bra{w_i, doc_j}}{\Tr(\sum_j \ket{w_i, doc_j}\bra{w_i, doc_j})}
\label{eq:ldop}
\end{gather}
As the document kets are superpositions of contexts, summing over their outer products in this way causes $\widehat{w_i}$ to have its own eigen base. If two documents use $w_i$ in the same sense, then their information will cluster in the same eigen ket of $\widehat{w_i}$. Otherwise their distributional information will likely end up in different eigen kets.

The division by the trace in Equation \ref{eq:ldop} causes $\widehat{w_i}$ to be normalized. Hence its eigen values sum to 1 and can be interpreted as probabilities. The spectrum of $\widehat{w_i}$ gives us a probability distribution over kets, each of which encodes a sense of the word $w_i$. This is a powerful data structure, as each sense of $w_i$ is not represented by a symbol, but rather by a multipartite ket which can be analyzed even further.

Even though the other models mentioned above differ from that of \cite{blacoe-kashefi-lapata:2013:NAACL-HLT} in several ways, they share a common idea: A word's meaning is represented by a density operator, the spectral decomposition of which reveals its senses and their relative frequency. In the following subsections we will put this idea into the context of quantum information theory and explain the application of the measures covered in Section \ref{sec:general} to the semantics of words.

In the following subsections we consider the example words \textit{book} and \textit{schedule} as an illustration of our information-theoretical analysis. Both words can be used as nouns or as verbs. \textit{book} as a noun has at least two senses: (1) the \textit{book} of Isaiah, (2) a collection of playing cards is a \textit{book} according to the game's rules. When used as a verb, \textit{book} can mean (1) to \textit{book} a venue or a ticket, or (2) to \textit{book} someone for a crime they committed. The noun \textit{schedule} is a calendar or planner. The verb \textit{schedule} is closely related, as it is the action performed by recording or planning a time and a place for some activity.



\subsection{Ambiguity}

Natural language is riddled with ambiguity. Words can have many kinds of ambiguity, including polysemy and homonymy. The former is the case when a word's senses are related. The latter results from a development of two unrelated words over time ending up with the same written and/or spoken form. In many dictionaries a word's various senses have separate entries specifying their part of speech and meaning in context. One measure of ambiguity is, therefore, the amount of dictionary entries a word has. Dictionaries often only list the most prevalent senses, though. Furthermore, if one sense is less common than another it should weigh less into the word's degree of ambiguity. This concern is naturally addressed by the entropy measure. With the entropy function, each sense's information content is weighted by it's probability and summed over. In our example we did not assign probabilities to the senses of \textit{book} and \textit{schedule}, but intuitively we can assume that the former has a higher entropy because it has more senses than the latter, each sense having a considerable probability mass.

\cite{Robin2015} use entropy to compute the ambiguity of words and phrases. They show through examples of adjective-noun modification and through nouns modified by relative clauses that this compositional process reduces the noun's entropy. In the same vein \cite{Blacoe:QI2014} demonstrated that the entropy of a sentence is generally lower than that of its words.

\subsection{Language Understanding}

When we hear a word in isolation its intended meaning could be any of its senses. However, this rarely happens in human communication. Other words serve to disambiguate it, especially those heard previously to the target word. The more recently another word was heard, the stronger is its potential effect on coming words. This is a type of priming. That is, recent words build an expectation for following words. They build the context in which the next words will be evaluated. If, for example, \textit{schedule} was heard shortly before the utterance of \textit{book}, then it (and potentially other intervening words) would have an effect on how the hearer decides which sense of \textit{book} was intended by the speaker.

This incremental composition can be cast in terms of measurement as in Section \ref{sec:measurement}. The previous words including \textit{schedule} constitute the state of the current system, that is our knowledge of the speaker's intended meaning. It is a mixed state, given the different senses of \textit{schedule}. When the hearer encounters the word \textit{book}, she compares all of its senses to the existing state of knowledge. Each sense ket of \textit{book} is assigned a probability through this comparison. We leave as an open question whether measuring the \textit{book} observable collapses the system's state to one of \textit{book}'s eigen kets or not during a cognitive process such as this (see Section \ref{sec:measurement}).


\subsection{Similarity}

What makes two words synonymous? Most words have multiple senses, even if the only differ through some semantic or syntactic feature. In order for two words to be synonyms, they would have to be interchangeable in every possible context. That is, they would have to have the same inventory of senses. When considering the disambiguation and composition of word meaning as above, an additional criterion is that the probabilities for each sense would have to be the exact same in both words.

It is virtually impossible to find two words that have all senses in common and with equal probability. We therefore need a similarity measure that takes differences in both of these two dimensions into account. Quantum fidelity is sensitive to both and renders a useful similarity value between 0 and 1 \cite{Esma2014}. The fidelity function can figure out which senses of one word align to which degree with which senses of another word. A sense of one word may overlap somewhat with several senses of another word. In the case of our example words, the verbal sense of \textit{schedule} overlaps highly with the first verbal sense of \textit{book}. All other pairs of sense kets will have very little overlap at most, thus preventing the corresponding probabilities from contributing to the overall fidelity.


\subsection{Entailment}

A word $w_1$ entails another word $w_2$ if $w_1$ is a special case of $w_2$. This relation becomes apparent in distributional semantics when $w_1$'s typical contexts are a subset of $w_2$'s typical contexts. In other words, $w_1$ is always replaceable by $w_2$, but not necessarily vice versa. This intuition has been captured by varying measures in the semantics literature \cite{Kotlerman:2010}. Additionally, relative entropy among quantum states has been suggested as an assymetrical similarity measure among words \cite{Esma2014}.
The relative entropy $S(\widehat{w_2}||\widehat{w_1})$ can be thought of as the amount of information that $w_1$ lacks in order to cover as wide a meaning spectrum as $w_2$.


\subsection{Correlation}

In our examplary creation of lexical density operators (Section \ref{sec:example}) we mentioned one way of making use of a multipartite Hilbert system: Each subsystem $V_{R_i}$ represents a type of syntactic relation $R_i$. This means that $\widehat{w}$'s eigen kets represent neighborhoods of context words around the word $w$. In \cite{Blacoe:QI2014} we are shown that these eigen kets contain correlation among their subsystems. For example, the two verbal senses of \textit{book} have different typical neighbors: ``The \textit{customer} booked a \textit{ticket}'' vs. ``The \textit{police} booked a \textit{thief}''. The correlation inside $\widehat{book}$ is among its arguments. \textit{customer} is correlated with \textit{ticket}, and \textit{police} is correlated with \textit{thief}. There would be no correlation among the subject and object subsystems if, for example, it were common and unambiguous to say ``The customer booked a thief'' and ``The police booked a ticket''.

The correlation measure used in \cite{Blacoe:QI2014} does not analyze how much of the found correlation is classical and how much is quantum. The \textit{book} example here shows that subjects and objects are not independent. However, this example only exhibits classical correlation. Since the pairs \textit{customer-ticket} and \textit{police-thief} are in different senses of \textit{book} this phenomenon can be modelled with a separable density operator.

Non-seperability comes with there being correlation inside at least one eigen ket. This is an additional piece of structure that may well find linguistic application. But it will take further investigation to identify specific ways of exploiting this advantage.



%% file: discussion.tex
\section{Conclusions}

We have surveyed several measures from probability and information theory and their quantum analogues.
While the classical versions of these measures apply to regular probability distributions, their quantum versions are for density operators. Measurement, fidelity and relative entropy are functions over (at least) two density operators $A, B$ which describe their interaction. We found that a common theme among these functions is the term that takes the inner product of all pairs of $A$'s and $B$'s eigen kets into account. Thereby all possible pairs of (eigen) probability values from $A$ and $B$ are able to effect the process outcome, whereas in classical distributions only those probabilities aligned one-to-one by an index are processed together.

This ``soft alignment'' of probabilities is potentially advantageous for any application model involving uncertainty and the interaction of distributions. We gave some intuitions as to how the field of distributional semantics may benefit from quantum information theory. We identified several existing works which already take advantage of density matrices and made suggestions about further use thereof. In particular the generalization of information-theoretic measures brings with it added possibilities for modeling complex interactions among words and phrases, or whatever objects are represented by density operators.

%

%% file: paper.bbl
\begin{thebibliography}{10}

\bibitem{Neumann1927}
Neumann, J.v.:
\newblock {Wahrscheinlichkeitstheoretischer Aufbau der Quantenmechanik}.
\newblock {Nachrichten von der Gesellschaft der Wissenschaften zu G\"ottingen,
  Mathematisch-Physikalische Klasse} \textbf{1927} (1927)  245--272

\bibitem{firth1957}
Firth, J.R.:
\newblock A synopsis of linguistic theory 1930-1955.
\newblock In: Studies in Linguistic Analysis.
\newblock Philological Society, Oxford (1957)  1--32

\bibitem{Turney:Pantel:2010}
Turney, P.D., Pantel, P.:
\newblock From frequency to meaning: Vector space models of semantics.
\newblock Journal of Artificial Intelligence Research \textbf{37} (2010)
  141--188

\bibitem{Erk:Survey:2012}
Erk, K.:
\newblock Vector space models of word meaning and phrase meaning: A survey.
\newblock Language and Linguistics Compass \textbf{6} (2012)  635--653

\bibitem{Sordoni2014}
Sordoni, A., Nie, J.Y.:
\newblock Looking at vector space and language models for ir using density
  matrices.
\newblock In Atmanspacher, H., Haven, E., Kitto, K., Raine, D., eds.: Quantum
  Interaction. Volume 8369 of Lecture Notes in Computer Science.
\newblock Springer Berlin Heidelberg (2014)  147--159

\bibitem{Esma2014}
Balkir, E.:
\newblock Using density matrices in a compositional distributional model of
  meaning.
\newblock Master's Thesis, University of Oxford (2014)

\bibitem{Robin2015}
Piedeleu, R., Kartsaklis, D., Coecke, B., Sadrzadeh, M.:
\newblock Open system categorical quantum semantics in natural language
  processing.
\newblock arXiv:1502.00831 (2015)

\bibitem{Coecke2010}
Coecke, B., Sadrzadeh, M., Clark, S.:
\newblock Mathematical foundations for a compositional distributional model of
  meaning. lambek festschrift.
\newblock Linguistic Analysis \textbf{36} (2010)  345--384

\bibitem{blacoe-kashefi-lapata:2013:NAACL-HLT}
Blacoe, W., Kashefi, E., Lapata, M.:
\newblock A quantum-theoretic approach to distributional semantics.
\newblock In: Proceedings of the 2013 Conference of the North American Chapter
  of the Association for Computational Linguistics: Human Language Technology,
  Atlanta, GA (2013)  847--857

\bibitem{Nielsen:Chuang:10}
Nielsen, M.A., Chuang, I.L.:
\newblock Quantum Computation and Information Theory.
\newblock Cambridge University Press, Cambridge (2010)

\bibitem{vonNeumann:1932}
von Neumann, J.:
\newblock {{Mathematische Grundlagen der Quantenmechanik}}.
\newblock Springer, Berlin (1932)

\bibitem{Vedral:RelativeEntropy}
Vedral, V.:
\newblock The role of relative entropy in quantum information theory.
\newblock arXiv:quant-ph/0102094 (2001)

\bibitem{Jensen:Inequality}
Jensen, J.:
\newblock Sur les fonctions convexes et les in\'egalit\'es entre les valeurs
  moyennes.
\newblock Acta Mathematica \textbf{30} (1906)  175--193

\bibitem{Vedral:discord}
Modi, K., Brodutch, A., Cable, H., Paterek, T., Vedral, V.:
\newblock The classical-quantum boundary for correlations: Discord and related
  measures.
\newblock Rev. Mod. Phys. \textbf{84} (2012)  1655--1707

\bibitem{Werner1989}
Werner, R.F.:
\newblock Quantum states with einstein-podolsky-rosen correlations admitting a
  hidden-variable model.
\newblock Phys. Rev. A \textbf{40} (1989)  4277--4281

\bibitem{Vedral:Correlation}
Henderson, L., Vedral, V.:
\newblock Classical, quantum and total correlations.
\newblock arXiv:quant-ph/0105028 (2001)

\bibitem{Blacoe:QI2014}
Blacoe, W.:
\newblock Semantic composition inspired by quantum measurement.
\newblock In Atmanspacher, H., Bergomi, C., Filk, T., Kitto, K., eds.: Quantum
  Interaction. Lecture Notes in Computer Science.
\newblock Springer International Publishing (2015)  41--53

\bibitem{Kotlerman:2010}
Kotlerman, L., Dagan, I., Szpektor, I., Zhitomirsky-Geffet, M.:
\newblock Directional distributional similarity for lexical inference.
\newblock Special Issue of Natural Language Engineering on Distributional
  Lexical Semantics \textbf{16} (2010)  359--389

\end{thebibliography}
